\documentclass{eas}
\usepackage[dvipsnames,usenames]{color}
\usepackage{graphicx}
\usepackage{hyperref}

\begin{document}

\title{Nearby Low-Mass Hypervelocity Stars}
\runningtitle{Nearby Hypervelocity Stars}
\author{Yanqiong Zhang}\address{Shanghai Astronomical Observatory;
zhangyq@shao.ac.cn, msmith@shao.ac.cn}
\author{Martin C. Smith$^1$}
\author{Jeffrey L. Carlin}\address{Rensselaer Polytechnic Institute, Troy \&
  Earlham College, Richmond, USA}
%
\begin{abstract}
Hypervelocity stars are those that have speeds exceeding the
escape speed and are hence unbound from the Milky Way. We investigate
a sample of low-mass hypervelocity candidates obtained using data from
the high-precision SDSS Stripe 82 catalogue, which we have combined
with spectroscopy from the 200-inch Hale Telescope at Palomar 
Observatory. We find four good candidates, but without metallicities
it is difficult to pin-down their distances and therefore total
velocities. Our best candidate has a significant likelihood that it is
escaping the Milky Way for a wide-range of metallicities.
\end{abstract}
\maketitle

\section{Introduction}

The peculiar velocities of traditional `runaway' stars are typically
greater than 30 km/s. In the solar neighbourhood approximately $10 \%$
-- $30 \%$ of O-type stars and $5 \%$ -- $10 \%$ of B-type stars are
runaways (Gies \cite{Gies87}, Stone \cite{Stone91}). Most runaway 
stars are ejected from the Galactic plane shortly after their formation
(Ramspeck {\em et al.\/} \cite{Ramspeck01}).
In 2005 a late B-type star was discovered in the outer halo
having a heliocentric radial velocity of 853 $\pm$ 12 km/s, which is
significantly in excess of the Galactic escape speed at that location
(Brown {\em et al.\/} \cite{Brown05}). Such stars are called `hypervelocity stars' (HVS).
Hills (\cite{Hills88}) predicted the existence of HVSs, suggesting that they can
form via tidal disruption of a stellar binary by the central
supermassive black hole of the Milky Way. In this scenario one of the
binary stars is captured by the black hole, while the other is
ejected, often at an extremely high velocity. Currently around 27
A/B/O-type unbound HVSs have been identified in recent years
(for example, Brown {\em et al.\/} \cite{Brown14} and references therein, Zheng {\em et al.\/} \cite{Zheng14}).
Other low-mass F/G/M-type HVS candidates have been claimed
(Li {\em et al.\/} \cite{Li12}, Palladino {\em et al.\/} \cite{Palladino14}, Zhong {\em et al.\/} \cite{Zhong14}), but most of these 
studies have found that their HVS candidates are not consistent with
Galactic centre ejection.
In this work we use the Sloan Digital Sky Survey's Stripe 82
data to identify new low-mass hypervelocity star candidates.

\section{Data and methods}
Stripe 82 is a 250 deg$^2$ field along the celestial equator in the
south Galactic cap, which has been repeatedly imaged by SDSS over 7
years allowing for studies of the variable sky and identification of
many kinds of transient phenomena 
(e.g. Sesar {\em et al.\/} \cite{Sesar07}, Watkins {\em et al.\/} \cite{Watkins09}).
The Stripe 82 dataset has also been used to construct high accuracy
proper motion catalogues (Bramich {\em et al.\/} \cite{Bramich08}, Koposov {\em et al.\/} \cite{Koposov13}). These
catalogues have been exploited to hunt for white dwarfs
(Vidrih {\em et al.\/} \cite{Vidrih07}) and analyse the kinematic properties of Galactic
disc and halo stars (Smith {\em et al.\/} \cite{Smith09a},
\cite{Smith09b}, \cite{Smith12}).

\begin{figure}
  \begin{center}
    \includegraphics[width=0.49\linewidth]{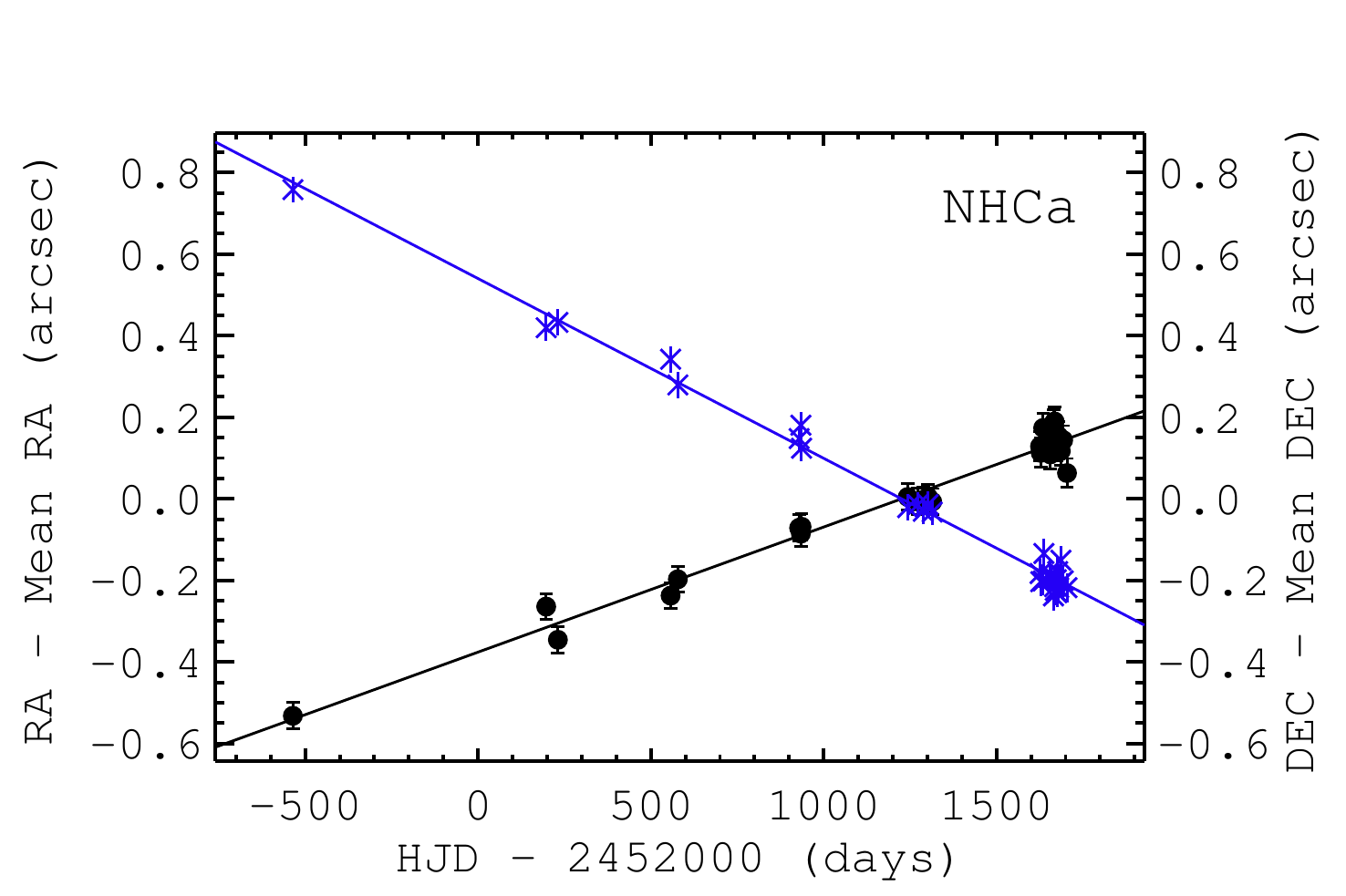} 
    \hfill  
    \includegraphics[width=0.49\linewidth]{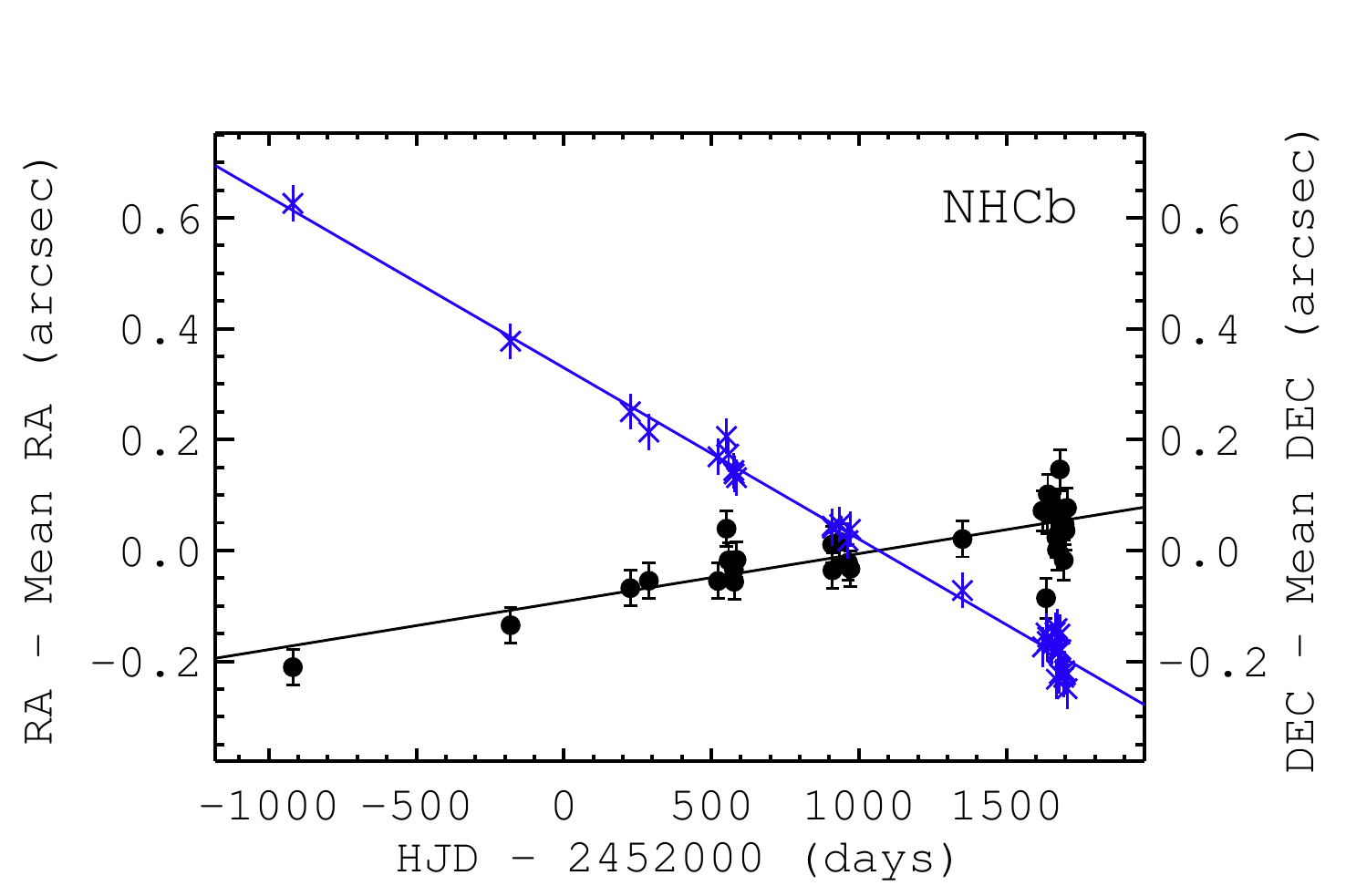} 
    \hfill  
    \includegraphics[width=0.49\linewidth]{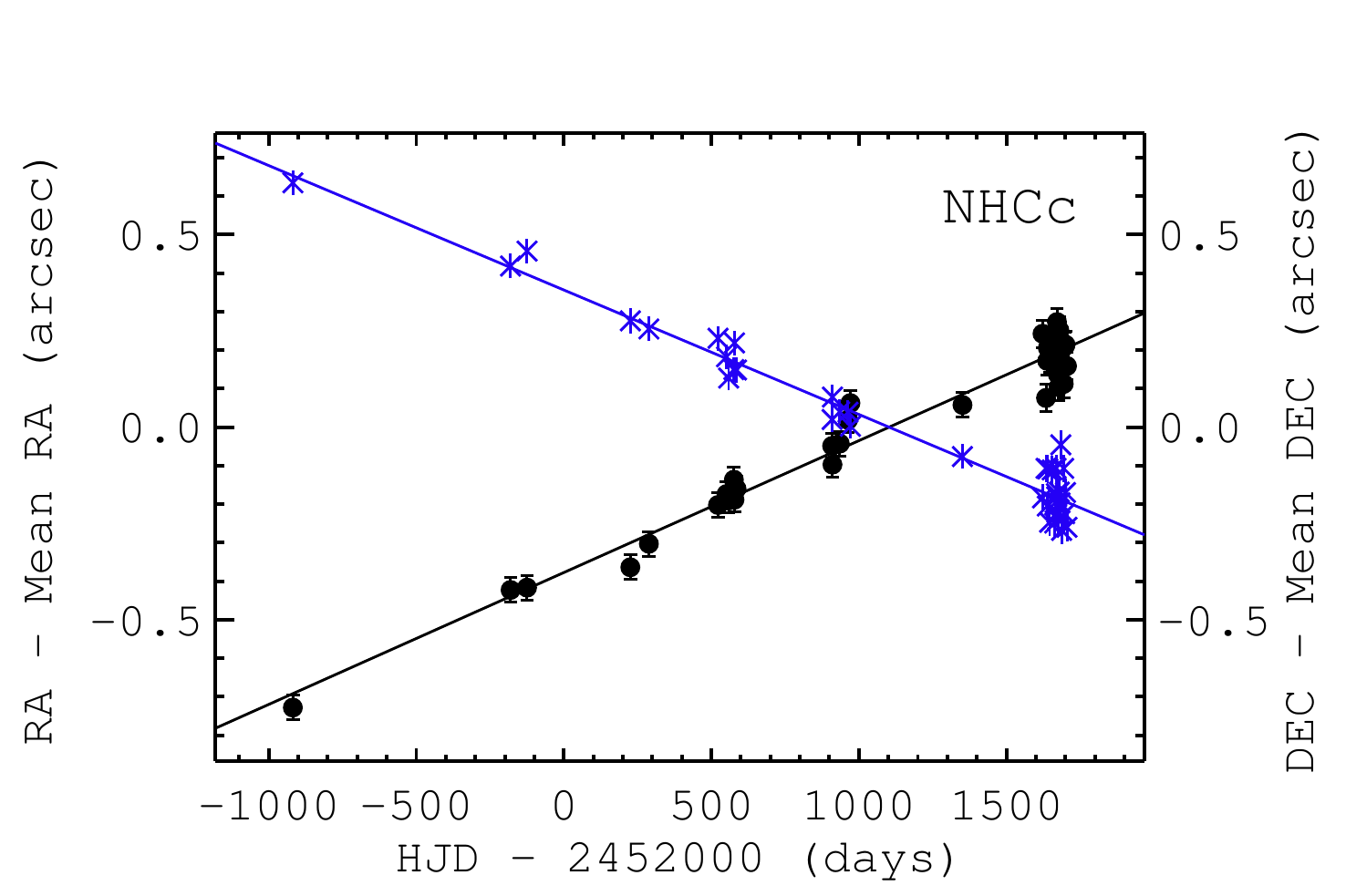} 
    \hfill  
    \includegraphics[width=0.49\linewidth]{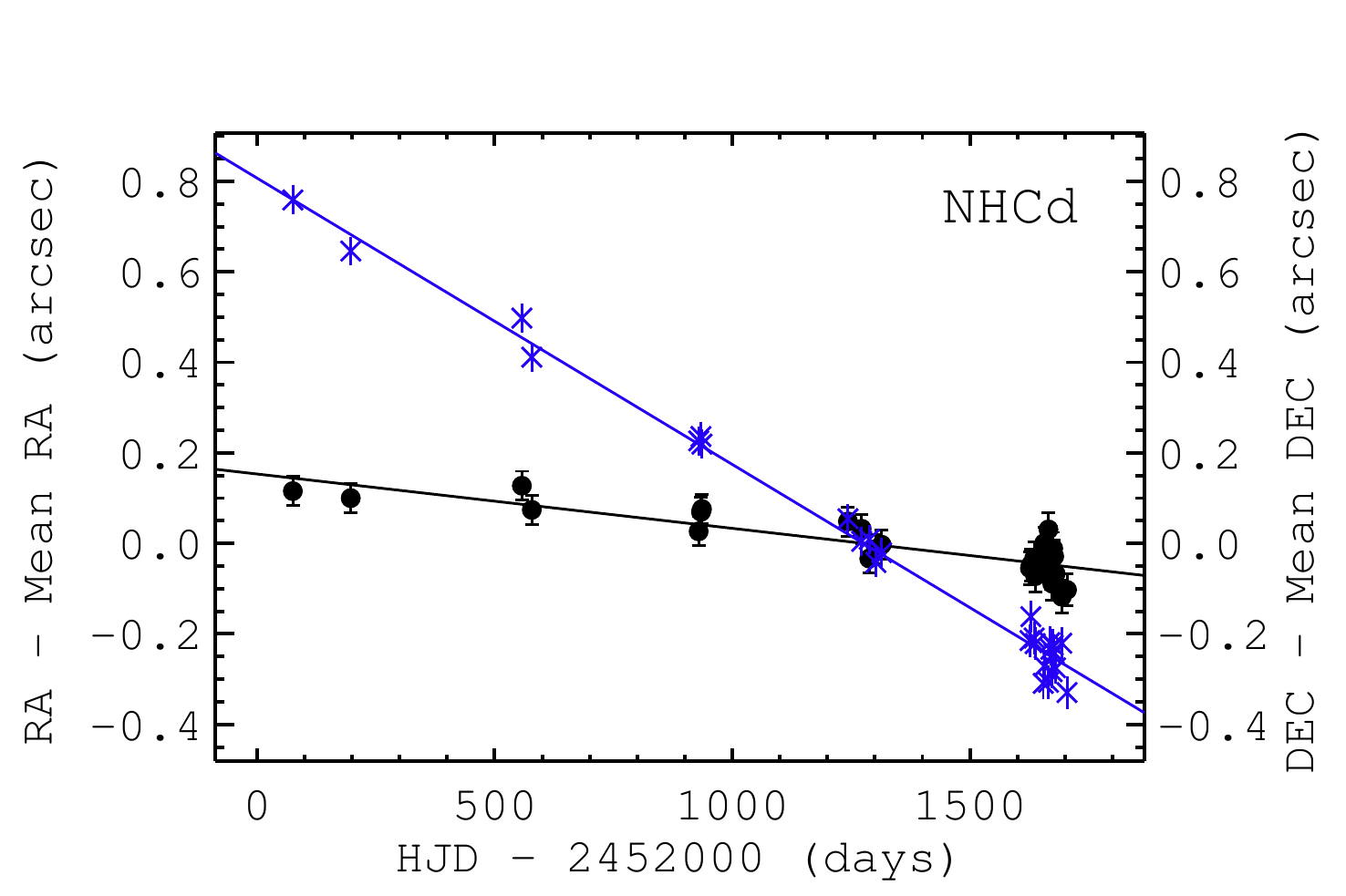}
    \caption{Measured proper motions of our candidates HVSs. The black  
      circles represent the proper motion in RA and the blue crosses in Declination.}
  \label{fig:propermotion}
  \end{center}
\end{figure}

Using this dataset we have attempted to find nearby hypervelocity
star candidates.
We first de-reddened all photometry using the
maps of Schlegel {\em et al.\/} (\cite{Schlegel98}).
We then identify red main-sequence stars following
Bond {\em et al.\/} \cite{Bond10}: $0.6 < g - r < 1.6$, $-0.15 < -0.27r + 0.8i - 0.534z +
0.054 <0.15$.
However, since the majority of very high proper motion stars are
nearby white dwarfs, we exclude these by removing all objects
bluer than $g - r = 0.9$.
In order to estimate a distance (and hence tangential velocity) we
need to know the metallicity of these stars. Although reliable
photometric metallicity estimators exist (e.g. Ivezi\'c {\em et al.\/} \cite{Ivezic08}), these
are not applicable to K/M stars. Therefore we have conservatively
assumed at this stage that their metallicity is -2 dex, which provides
a conservative estimate on the tangential velocity (see Section
\ref{sec:results}). We adopt the photometric distance relation of 
{Ivezi\'c \em et al.\/} \cite{Ivezic08} (including the turn-off correction of
{Smith \em et al.\/} \cite{Smith09b}). We then choose all stars with heliocentric
tangential velocity larger than 400 km/s and proper motion errors 
less than 1 mas/yr. This results in a sample of 11 HVSs candidates. We
check the proper motions of our candidate HVSs using the data from
(Bramich {\em et al.\/} \cite{Bramich08}), as can be seen in Figure
\ref{fig:propermotion}. Clearly our four best candidates are 
fast moving, meaning that there are no spurious high proper motion
stars. This is often a problem for candidate HVSs obtained using
catalogues constructed from cross-matching different surveys (see
Vickers {\em et al.\/} \cite{Vickers15}).

In order to determine the full 3D velocities we needed spectroscopy of
these 11 candidates, which we obtained using Double Spectrograph on
the Hale 200-inch telescope at Palomar Observatory. Our resolution was
around R $\sim$ 2000, giving an internal accuracy of a few km/s.


\section{Candidate hypervelocity stars}
\label{sec:results}

\begin{figure}
  \begin{center}
    \includegraphics[width=0.6\linewidth]{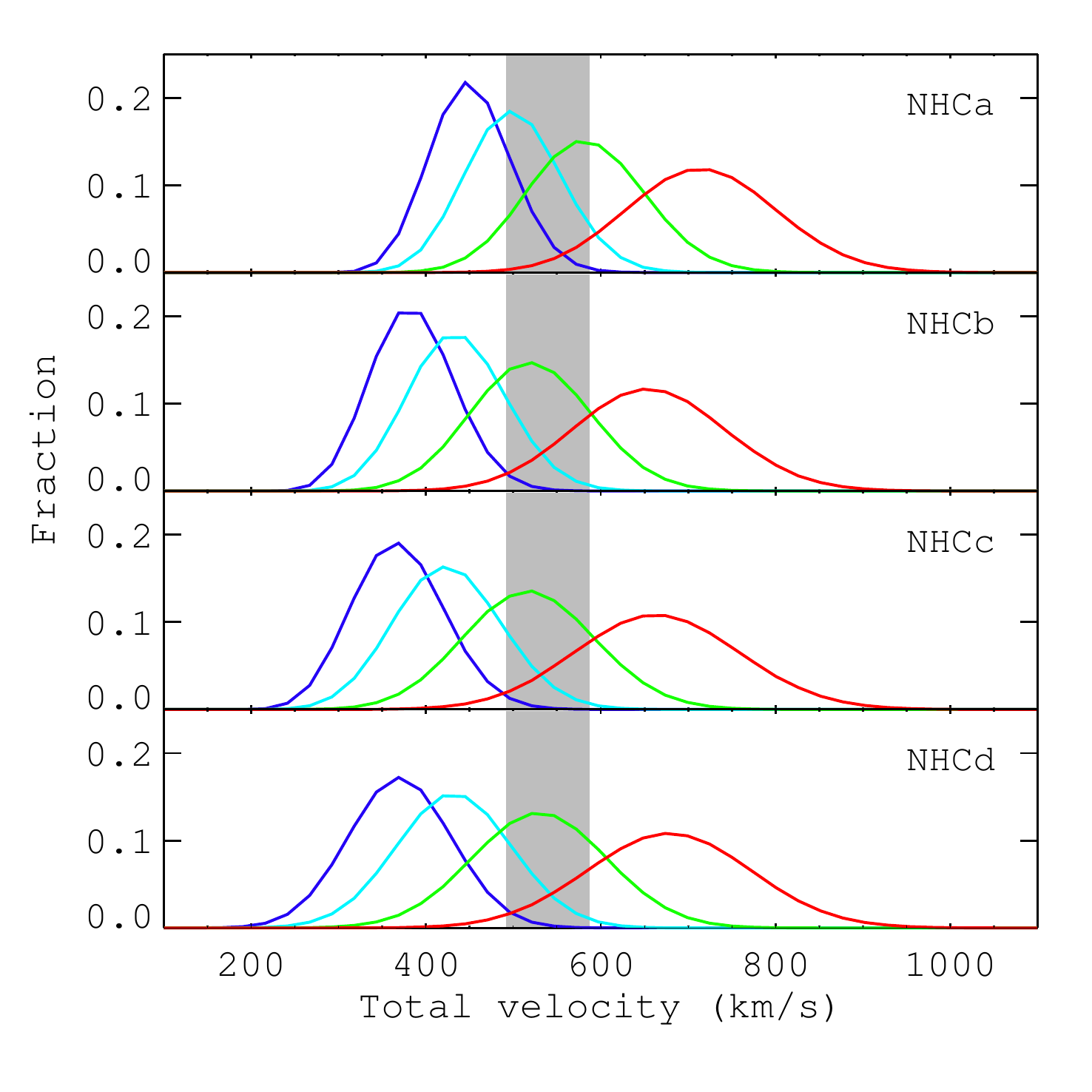} 
    \caption{The probability distribution of total velocity (in
      the Galactic rest frame) for our best candidates. The vertical
      grey stripe shows the local escape velocity
      (Piffl {\em et al.\/} \cite{Piffl14}). From left-to-right the curves correspond to
      distances calculated assuming [Fe/H] = -2, -1.5, -1, -0.5 dex.}
    \label{fig:escape}
  \end{center}
\end{figure}

\begin{table}
  \begin{center}
    \caption{Nearby hypervelocity star candidates}
    \label{tab:kinematics}
    \begin{tabular}{lccccc}
      \hline
      ID & $\mu_\alpha cos \delta $ &
      $\mu_\delta$  & $D^{\ast}$ &
      $v_{\rm r}$ & $v_{\rm tot}^{\ast,\dagger}$\\
      & $(\rm{\mu mas/yr}) $ & $(\rm{\mu mas/yr})$ & (kpc) &
      $(\rm{km/s})$& $(\rm {km/s})$
      \\  
      \hline      
      NHCa & 99.92 $\pm$ 0.65 & -166.12 $\pm$ 0.65 & 0.6  &
      -439.2 $\pm$ 1.4 & 501 $\pm$ 54\\   

      NHCb & 27.52 $\pm$ 0.57 & -108.89 $\pm$ 0.57 & 1.1 &
     -235.4 $\pm$ 10 & 436 $\pm$ 55\\  

      NHCc & 120.19 $\pm$ 0.70 & -122.11 $\pm$ 0.70 & 0.8 &
     186.2 $\pm$ 5.4 & 426 $\pm$ 61 \\   

      NHCd & -45.24 $\pm$ 0.96 &  -225.15 $\pm$ 0.96 & 0.6 &
    -123.4 $\pm$ 5.2 & 432 $\pm$ 66\\  
      \hline      
    \end{tabular}
  \end{center}
  \emph{\rm{$^{\ast}$ Distance calculated assuming [Fe/H] = -1.5 dex.}}
  ~~~\emph{\rm{$^{\dagger}$ In the Galactic rest frame.}}
\end{table}
For these 11 candidates we calculate their full 3D positions and
velocities, taking the Solar peculiar velocity from
Sch{\"o}nrich {\em et al.\/} (\cite{Schonrich12}). Our four
fastest-moving stars, which we label Nearby Hypervelocity Candidates
(NHC), are presented in Table \ref{tab:kinematics}.
Figure \ref{fig:escape} shows the probability distribution function
for the total velocity (with respect to the Galactic rest frame) for
our four best candidates. This calculation folds in all uncertainties,
e.g. distance, proper motion, radial velocity, Solar motion. The
distance is calculated using the relation from Ivezi\'c {\em et al.\/}
(\cite{Ivezic08}); since this relation depends on metallicity, which
for our stars is currently unknown, we have assumed four values of
${\rm [Fe/H]}$. If the metallicity is known, then the uncertainty on
the distance is around 10--15\% (Ivezi\'c {\em et al.\/}
\cite{Ivezic08}, Smith {\em et al.\/} \cite{Smith09b}).
The vertical band shows the 90\% confidence interval on the local
escape speed ($533^{+54}_{-41}$ km/s; Piffl {\em et al.\/} \cite{Piffl14}). 
As expected, the total velocity decreases as the metallicity
decreases. This is because the absolute magnitude (and hence distance
and tangential velocity) of a star is correlated to its metallicity,
with lower-metallicities leading to fainter absolute magnitudes (and
smaller distances).
Table \ref{tab:escape} shows the probabilities that our candidates
have speed exceeding the local escape speed. As these are most-likely
halo stars, with metallicities typically around $-1$ to $-2$ dex,
all four candidates have significant probabilities that they are not
bound to the Milky Way. The fastest moving star, NHCa, is likely to be
escaping even if the metallicity is as low as $\sim -1.5$ dex.

\begin{table}
  \begin{center}
    \caption{The probability that our candidates are unbound from the Milky Way.}
    \label{tab:escape}
    \begin{tabular}{lcccc}
      \hline
      ID & \multicolumn{4}{c}{Probability$\left(v_{\rm tot} > v_{\rm esc}\right)$}\\
      & [Fe/H] = - 2.0 & [Fe/H] = - 1.5 &
      [Fe/H] = - 1.0&[Fe/H] = - 0.5
      \\  
      \hline      
      NHCa & 4.3 \% & 27.5 \% & 77.8 \%  &  98.7 \%\\  
      NHCb & 0.2 \% & 4.5 \% & 42.7 \%  &  92.5 \%\\   
      NHCc & 0.2 \% & 4.4 \% & 42.1 \%  &  92.4 \%\\   
      NHCd & 0.3 \% & 6.3 \% & 48.3 \%  &  94.1 \%\\   
      \hline      
    \end{tabular}
  \end{center}
 
\end{table}


In conclusion, we have presented a sample of candidate hypervelocity
stars that may be unbound from the Milky Way.
Accurate proper motions and distances are crucial to such analyses. We
have overcome the former using the reliable, high-precision Stripe 82
catalogue (Bramich {\em et al.\/} \cite{Bramich08}). We are less certain on the latter, as we
need to know the metallicity of our candidates before we can pin down
their distances. We are now working on estimating metallicity from our
spectra and will report our findings in a future work. Clearly Gaia
will revolutionise our understanding of the population of unbound
stars in the Milky Way and studies like ours show the potential from
even very small samples (in our case only 250 deg$^2$). Curiously,
all of our four best candidates are infalling, which means that none
are consistent with Galactic centre ejection. Currently the origin of
low mass HVSs is controversial and this is something we aim to address
in future work.


{\let\thefootnote\relax\footnote{
The authors wish to thank Alberto Rebassa-Mansergas, Matthew Molloy
and Eric Peng for assistance.
This work is supported by the CAS One Hundred Talent Fund, NSFC Grants  
11173002 \& 11333003, and the Gaia Research for European Astronomy
Training (GREAT-ITN) Marie Curie network, funded through the European
Union Seventh Framework Programme (FP7/2007-2013) under grant
agreement no 264895. It has used data obtained through the Telescope
Access Program (TAP), which has been funded by the Strategic Priority
Research Program "The Emergence of Cosmological Structures" (Grant
No. XDB09000000), National Astronomical Observatories, Chinese Academy
of Sciences, and the Special Fund for Astronomy from the Ministry of
Finance. Observations obtained with the Hale Telescope at Palomar
Observatory were obtained as part of an agreement between the National
Astronomical Observatories, Chinese Academy of Sciences, and the
California Institute of Technology.}}


\end{document}